\newcommand{\odds}{\ensuremath{\mathcal{O}}}
\documentclass{aa}
\usepackage{times}
\begin{document}

   \thesaurus{03     
              (13.07.01; 
               08.19.4)} 
\title{Evidence Against an Association Between Gamma-Ray Bursts and
Type I Supernovae}

\author{C. Graziani 
   \and D. Q. Lamb
   \and G. H. Marion}
\institute{Department of Astronomy and Astrophysics, University
of Chicago, \\ 5640 South Ellis Avenue, Chicago, IL 60637}

\date{Received December 15, 1998}

\maketitle

\begin{abstract}
We use a rigorous method, based on Bayesian inference, for
calculating the odds favoring the hypothesis that any particular class
of astronomical transients produce gamma-ray bursts over the hypothesis
that they do not.  We then apply this method to a sample of 83 Type Ia 
supernovae and a sample of 20 Type Ib-Ic supernovae.  We find
overwhelming odds against the hypothesis that all Type Ia supernovae
produce gamma-ray bursts, whether at low redshift ($10^{9}:1$)  or
high-redshift ($10^{12}:1$), and very large odds ($6000:1$) against the
hypothesis that all Type Ib, Ib/c, and Ic supernovae produce observable
gamma-ray bursts.  We find large odds ($34:1$) against the hypothesis
that a fraction of Type Ia supernovae produce observable gamma-ray
bursts, and moderate odds ($6:1$) against the hypothesis that a
fraction of Type Ib-Ic supernovae produce observable bursts.

\end{abstract}

\section{Introduction}

The discovery of supernova 1998bw (Galama et al. \cite{galama98})
within the 8 arcminute radius of the BepppoSAX WFC error circle for
GRB 980425 (Soffitta et al. \cite{soffitta98}) has led to the hypothesis
that some GRB sources are Type Ib-Ic SNe.  

There are some serious difficulties with this interpretation of the
data for GRB 980425/SN 1998bw:  the supernova occurred outside the NFI
error circle of a fading X-ray source (Pian et al.
\cite{pian98a},\cite{pian98b}, \cite{pian99} ; Piro et
al.\cite{piro98}).  This source had a temporal decay consistent with a
power-law index of $\sim 1.2$ (Pian et al. \cite{pian98b}), which
resembles the temporal behavior of X-ray afterglows seen in almost
every other GRB followed up with the SAX NFI.  It must therefore be
viewed as a strong candidate to be the X-ray afterglow of GRB 980425. 

Moreover, if the association between GRB 980425 and SN 1998bw were
true, the luminosity of this burst would be $\sim 10^{46}$ erg s$^{-1}$
and its energy  would be $\sim 10^{47}$ erg.  Each would therefore be
five orders of magnitude less than that of other bursts, and the
behavior of the X-ray and optical afterglow would be very different
from those of the other BeppoSAX bursts, yet the burst itself is
indistinguishable from other BeppoSAX and BATSE GRBs with respect to
duration, time history, spectral shape, peak flux, and and fluence
(Galama et al. \cite{galama98}).

In view of these difficulties, the safest procedure is to regard the
association as a hypothesis that is to be tested by searching for
correlations between SNe and GRB in catalogs of SNe and GRBs, excluding
SN 1998bw and GRB 980425. Wang \& Wheeler (\cite{wang98a}; see also
Wang \& Wheeler \cite{wang98b}) have  correlated BATSE GRB with Type Ia
and with Type Ib-Ic Sne.  They found that the data was ``consistent''
with the assumption of an association between GRB and Type Ib-Ic SNe.

In the present work we improve upon the Wang \& Wheeler correlative
study by introducing an analysis  method based on Bayesian inference,
and therefore using the likelihood function, that incorporates
information about the BATSE position errors in a non-arbitrary way and
that is free of the ambiguities of {\it a posteriori} statistics.  The
method also accounts the fact that the BATSE temporal exposure is less
than unity.

\section{Methodology}

We use the BATSE 4B GRB Catalog (Meegan et al. \cite{meegan98}), and
BATSE bursts that occurred subsequent to the 4B catalog but before 1
May 1998.  We also use the Ulysses supplement to the BATSE 4B catalog,
which contains 219 BATSE bursts for which 3rd IPN annuli have been
determined (Hurley et al. \cite{hurley98}).  Hurley (private
communication, 1998) has kindly made available at our request 3rd IPN
annuli for an additional 9 BATSE bursts that occurred subsequent to the
period of the BATSE 4B catalog but before 1 May 1998.

We have compiled three Type I SNe samples.  The first is a sample of 37
Type Ia SNe at low redshift ($z < 0.1$) from the CfA SN Search Team
(Riess 1998, private communication).  The second is a sample of 46
moderate redshift ($0.1 < z < 0.830$) Type Ia SNe from the Supernova
Cosmology Project (Perlmutter 1998, private communication).  The third
sample consists of 20 Type Ib, Ib/c, and Ic SNe compiled from IAU
Circulars and various SNe catalogs.

We compare three hypotheses:

\noindent $H_1$: The association between SNe and GRBs is real.  If a SN
is observed, there is a chance $\epsilon$ that BATSE sees the
associated GRB, where $\epsilon$ is the average BATSE temporal
exposure.  While $\epsilon$ varies with Declination, the variation is
modest and we neglect it.  The probability density for the time of
occurrence of the $i$th supernova is assumed uniform in the interval
$[t_i,t_i+\tau_i]$, so that all GRBs that occur in that interval have
an equal prior probability of being associated with the SN.

\noindent $H_1^\prime$: The association is real, but only a fraction
$f$ of detectable SNe produce detectable GRB.

\noindent $H_2$: There is no association between SNe and GRBs.

We calculate the Bayesian Odds, $\odds$ favoring $H_1$ over $H_2$.  The
details of the calculation are presented in Graziani et al
(\cite{graziani98}).  We separately compare $H_1^\prime$ to $H_2$,
denoting the odds favoring $H_1^\prime$ over $H_2$ by $\odds^\prime$.
Finally, we calculate the posterior probability distribution for the
parameter $f$ in the model $H_1^\prime$, and infer an upper limit on
its value.

\section{Results}

We find overwhelming odds against the hypothesis that all Type Ia
supernovae produce gamma-ray bursts, whether at low redshift
($\odds=10^{9}:1$)  or high-redshift ($\odds=10^{12}:1$).   We also
find large odds ($\odds^\prime=34:1$) against the hypothesis that a
fraction of Type Ia supernovae produce observable gamma-ray bursts

We find very large odds ($\odds=6000:1$) against the hypothesis that
all Type Ib, Ib/c, and Ic supernovae produce observable gamma-ray
bursts.  We also find moderate odds ($\odds^\prime=6:1$) against the
hypothesis that a fraction of Type Ib-Ic supernovae produce observable
bursts.  If we nevertheless assume that this hypothesis is correct, we
find that the fraction $f_{\rm SN}$ of Type Ib, Ib/c and Ic SNe that
produce observable GRBs must be less than 0.17, 0.42, and 0.70 with
68\%, 95\%, and 99.6\% probability, respectively.  These limits are
relatively weak because of the modest size (20 events) of our sample of
Type Ib-Ic SNe.

\section{Discussion}

We find large odds against the hypothesis that all Type Ib, Ib/c, and
Ic supernovae produce observable gamma-ray bursts, the specific
hypothesis considered by Wang \& Wheeler (\cite{wang98a},
\cite{wang98b}).  The odds against the hypothesis that a fraction of
Type Ib-Ic supernovae produce observable bursts are more moderate,
because they account for the possibility that $f$ is so small that no
SN-produced GRBs are detected, in which case the data can not
distinguish between $H_1^\prime$ and $H_2$.

Type Ib, Ib/c and Ic SNe are now being found at a rate of about eight a
year, so that the size of the sample of known Type Ib-Ic SNe should
triple within about five years.  One might hope that future analyses,
using the statistical methodology that we have presented here, could
either show that the association between Type Ib-Ic SNe and GRBs is
rare, or confirm the proposed association.  Unfortunately, achieving
the former will be difficult: the limit on the fraction $f_{\rm SN}$ of
Type Ib-Ic SNe that produce observable GRBs scales like $N_{\rm
SN}^{-1}$ for large $N_{\rm SN}$, and therefore tripling the size of
the sample of known Type Ib-Ic SNe without observing an additional
possible SN -- GRB association would only reduce the 99.7\% probability
upper limit on $f_{\rm SN}$ to 0.24.

The parameter $f$ represents the fraction of observable SNe which can
produce observable GRBs.  Some observable SNe might not produce
observable GRBs because of intrinsic effects, such as beaming, whereas
others might not do so because the sampling distance for SN-produced
GRB could be less than the sampling distance for the SNe themselves. 
It is possible to separate these two effects, by writing $f=f_{\rm
intrinsic}\times f_{\rm sampling}$.  Obviously, the more interesting
quantity is $f_{\rm intrinsic}$, since it addresses the nature of the
GRB sources.  

We may attempt to find a constraint on $f_{\rm intrinsic}$ by assuming
the correctness of the identification of SN 1998bw with GRB 980425 and
using that association to estimate $f_{\rm sampling}$, under the
(dubious) assumption that the GRBs produced by Type Ib-Ic SNe are
standard candles.  The result is $f_{\rm sampling}\approx 1.9\times
10^{-3}$. This is rather bad news for the prospect of constraining
$f_{\rm intrinsic}$, since the product of $f_{\rm intrinsic}$ and
$f_{\rm sampling}$ is only constrained by the data to be less than
about 0.7.  Thus this argument can place no constraint on $f_{\rm
intrinsic}$.  Elsewhere, we show that placing a meaningful constraint
on $f_{\rm intrinsic}$ would require a GRB experiment approximately 80
times more sensitive than BATSE (Graziani et al. \cite{graziani98}).

One can also approach the proposed association between SNe and GRBs
from the opposite direction.   The interesting question, from this
point of view, is what fraction $f_{\rm GRB}$ of the GRBs detected by
BATSE could have been produced by Type Ib-Ic SNe?  (note that this is
different from the question addressed by Kippen et al.
[\cite{kippen98}], who constrained the fraction of BATSE GRB that could
have been produced by {\it known} SNe).  Such a limit may be derived
under the same assumptions as made in the previous paragraph. We find
that no more than $\sim 90$ SNe could have produced GRB detectable by
BATSE, indicating that $f_{\rm GRB}$ can be no more than about 5\%.

\end{document}